# Orbital Period Ratios and Fibonacci Numbers in Solar Planetary and Satellite Systems and in Exoplanetary Systems

Vladimir Pletser

Technology and Engineering Center for Space Utilization,
Chinese Academy of Sciences, Beijing, China
Vladimir.Pletser@csu.ac.cn

**Abstract**
It is shown that orbital period ratios of successive secondaries in the Solar planetary and giant satellite systems and in exoplanetary systems are preferentially closer to irreducible fractions formed with Fibonacci numbers between 1 and 8 than to other fractions, in a ratio of approximately 60% vs 40%. Furthermore, if sets of minor planets are chosen with gradually smaller inclinations and eccentricities, the proximity to Fibonacci fractions of their period ratios with Jupiter or Mars' period tends to increase. Finally, a simple model explains why the resonance of the form $\frac{P_1}{P_2} = \frac{p}{p+q}$, with $P_1$ and $P_2$ orbital periods of successive secondaries and $p$ and $q$ small integers, are stronger and more commonly observed.

**Keywords**: Orbital period ratios; Near mean motion resonances; Fibonacci numbers; Solar System; Giant planet satellite systems; Minor planets; Exoplanetary systems

## 1 Introduction

The discovery of the Trappist-1 system of seven planets (Gillon et al., 2017; Luger et al., 2017) with five out of six orbital period ratios being close to ratios of Fibonacci integers (Pletser and Basano, 2017) has prompted a search among other planetary and satellite systems of the Solar System and of exo-planetary systems to assess whether Fibonacci numbers intervene more often in integer fractions close to ratios of orbital periods. It is found that ratios of Fibonacci numbers outnumber significantly ratios formed with other integers, when limited to the most significant ratios of small integers between 1 and 8.

To recall, the sequence of Fibonacci integers is formed by integers equal to the sum of the two previous integers in the sequence, $F_i = F_{i-1} + F_{i-2}$ with $F_1 = 1$ and $F_2 = 1$ (or alternatively with $F_0 = 0$ and $F_1 = 1$), yielding the sequence 1, 1, 2, 3, 5, 8, 13, … A large number of formulas and equations exist involving Fibonacci numbers (see e.g. Chandra and Weisstein, 2017, and references therein).

Fibonacci numbers can be found in nearly all domains of Science, appearing when self-organization processes are at play and/or expressing minimum energy configurations. Several, but by far non-exhaustive, examples are given in (Pletser, 2017c) in biology, physics, astrophysics, chemistry and technology. In our Solar System, several attempts have been made to fit Fibonacci sequences to the orbital parameters of planets and satellite systems of the giant planets. Read (1970) claimed that the Fibonacci sequence can predict the distances of the moons of Jupiter,

Saturn and Uranus from their respective primary, and the planets have similarly a trend which follows the Fibonacci sequence with individual offsets attributed to planetary densities. This approach is hardly credible as the statistical treatment was poor and there was no physical explanation to support this claim. Recently, and although none of the actual planets are in exact resonance, Aschwanden (2018) has attributed to ratios of orbital periods of successive planets and giant planet satellites, harmonic ratios given by five dominant resonances, namely (3 : 2), (5 : 3), (2 : 1), (5 : 2), (3 : 1), and they conclude that the orbital period ratios tend to follow the quantized harmonic ratios in increasing order. One notices immediately that these five dominant resonances are all made of the second to the fifth Fibonacci numbers. Furthermore, Aschwanden and Scholkmann (2017) showed that the most prevailing harmonic ratios in 73% of 932 exoplanets pairs are (2:1), (3:2), and (5:3), which involve again the second to the fifth Fibonacci numbers.
We want in this paper to pursue this approach and to assess whether ratios of orbital periods of successive secondaries, planets, satellites of giant planets and exoplanets, follow a similar trend of being more often close to ratio of small Fibonacci integers.

## 2. Irreducible fractions with Fibonacci numbers

Considering only natural integers between 1 and 8, one can form only 22 irreducible fractions $\frac{n}{d}$ smaller than or equal to 1. Surprisingly, only ten are formed with the second to the sixth Fibonacci integers (1, 2, 3, 5, 8) and twelve are formed with only one of these five Fibonacci integers and another integer or with two non-Fibonacci integers (see Table 1).

TABLE 1: FRACTIONS WITH TWO FIBONACCI NUMBERS AND WITHOUT

| With two Fibonacci numbers | With one or no Fibonacci numbers |
|---|---|
| 1/1, 1/2, 1/3, 1/5, 1/8, 2/3, 2/5, 3/5, 3/8, 5/8 | 1/4, 1/6, 1/7, 2/7, 3/4, 3/7, 4/5, 4/7, 5/6, 5/7, 6/7, 7/8 |

Assuming a uniform distribution of ratios, each of these fractions should have a chance of 1/22 = 4.54 % of appearing and among these, irreducible fractions formed with Fibonacci numbers should appear with a total of 45.4 % chance. However, this over-simplistic approach should be modified to take into account the width of bins that include these fractions. Arranging these fractions $\frac{n_i}{d_i}$ by increasing order, we define 23 bins or "windows" around these fractions, as follows: for $1 \leq n_i < d_i \leq 8$, 22 bins from $\left[\frac{n_{i-1}}{d_{i-1}} + \left(\frac{\frac{n_i}{d_i} - \frac{n_{i-1}}{d_{i-1}}}{2}\right)\right]$ to $\left[\frac{n_i}{d_i} + \left(\frac{\frac{n_{i+1}}{d_{i+1}} - \frac{n_i}{d_i}}{2}\right)\right]$ around $\frac{n_i}{d_i}$ with $\frac{n_1}{d_1} = \frac{1}{8}$ and $\frac{n_0}{d_0} = \frac{1}{9}$, and a first window from 0 to $r_{inf} = \left[\frac{1}{9} + \left(\frac{\frac{1}{8} - \frac{1}{9}}{2}\right)\right] = 0.1181$. Figure 1 shows these windows and the location of the Fibonacci ratios and of other ratios.

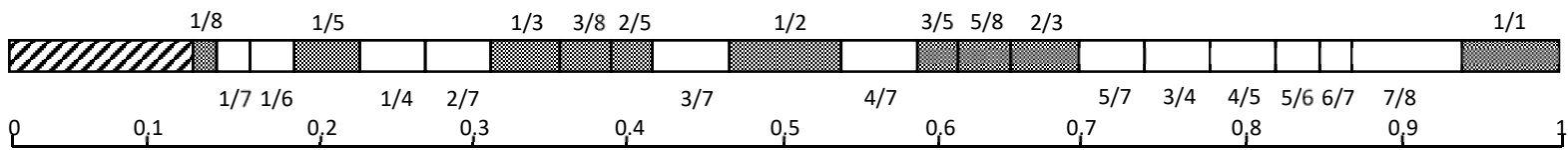

Figure 1: Location of the ten bins (dark) around fractions with Fibonacci numbers (top) and of the twelve bins (white) around fractions with only one or no Fibonacci numbers (bottom), and of the first window (hatched) between 0 and $r_{inf} = 0.1181$.

Assuming again a ratio uniform distribution, the chance of a real number between 0 and 1 to fall into one of these bins equals the width of the bin, i.e. $\frac{1}{2}\left(\frac{n_{i+1}}{d_{i+1}} - \frac{n_{i-1}}{d_{i-1}}\right)$, as shown in Table 2.

TABLE 2: CHANCE OF FALLING IN ONE OF THE BINS

| Fibonacci ratio bins | Chance | Other ratio bins | Chance |
|---|---|---|---|
| 1/8 | 1.59 % | 0 <...< 1/8 | 11.81 % |
| 1/5 | 4.17 % | 1/7 | 2.08 % |
| 1/3 | 4.46 % | 1/6 | 2.86 % |
| 3/8 | 3.33 % | 1/4 | 4.29 % |
| 2/5 | 2.68 % | 2/7 | 4.17 % |
| 1/2 | 7.14 % | 3/7 | 5.00 % |
| 3/5 | 2.68 % | 4/7 | 5.00 % |
| 5/8 | 3.33 % | 5/7 | 4.17 % |
| 2/3 | 4.46 % | 3/4 | 4.29 % |
| 1/1 | 6.25 % | 4/5 | 4.17 % |
|  |  | 5/6 | 2.86 % |
|  |  | 6/7 | 2.08 % |
|  |  | 7/8 | 7.14 % |
| **Total** | 40.10 % | **Total (without first bin)** | 48.10 % |
|  |  | **Total (with first bin)** | 59.90 % |

The largest bins are for 1/2 and 7/8 with width of 7.14% each and the smallest bin is for 1/8 with a width of 1.59%. The total chance of a number falling in one of the bins of Fibonacci fractions is 40.1%, while the chance of it falling in a bin with non-Fibonacci fractions is 59.9% or 48.1% if the first bin is counted or not.

### 3. Ratios of Orbital Periods

One considers ratios of orbital periods $p_i$ of successive prograde secondaries, planets or satellites, from the central primary outward, i.e. period ratios of non-consecutive or retrograde or distant secondaries that would yield too small period ratios (typically less than to $r_{inf} = 0.1181$) are not taken into account. We take the convention of expressing ratios $r_i$ of periods of the inner $(i -1)^{th}$ secondary to the outer $i^{th}$ secondary,

$$r_i = \frac{p_{i-1}}{p_i}$$
(1)

so as to obtain ratios between 0 and 1. Only neighbouring secondaries (planets or satellites) are considered in a first instance as stability of orbits in, or close to, resonance is ensured by gravitational interactions between close enough secondaries.

3.1 Solar planetary and giant planet satellite systems

We consider the eight planets of the Solar System plus Ceres and Pluto, and the main prograde satellites of the giant planets, having a diameter above 100 km. Only period ratios of successive neighbouring secondaries are considered. So, the resonances between Mimas and Tethys, between Enceladus and Dione, between large and smaller satellites and within rings are not counted in Table 3.

TABLE 3: SOLAR SYSTEM PLANETS AND SATELLITES

|  | Orbital Period (*) | Ratio $r_i = \frac{p_{i-1}}{p_i}$ | Closest Fraction |  | Orbital Period (*) | Ratio $r_i = \frac{p_{i-1}}{p_i}$ | Closest Fraction |
|---|---|---|---|---|---|---|---|
| **Sun** | | | | **Saturn** | | | |
| Mercury | 0.241 | -- | -- | Epimetheus | 0.694 | -- | -- |
| Venus | 0.615 | 0.3919 | 2/5 | Janus | 0.695 | 0.9995 | 1/1 |
| Earth | 1.000 | 0.6150 | 5/8 | Mimas | 0.942 | 0.7371 | 3/4 |
| Mars | 1.881 | 0.5316 | 1/2 | Enceladus | 1.370 | 0.6878 | 2/3 |
| Ceres | 4.601 | 0.4088 | 2/5 | Tethys | 1.888 | 0.7258 | 5/7 |
| Jupiter | 11.862 | 0.3879 | 2/5 | Dione | 2.737 | 0.6898 | 2/3 |
| Saturn | 29.457 | 0.4027 | 2/5 | Rhea | 4.518 | 0.6058 | 3/5 |
| Uranus | 84.018 | 0.3506 | 1/3 | Titan | 15.945 | 0.2834 | 2/7 |
| Neptune | 164.780 | 0.5099 | 1/2 | Hyperion | 21.277 | 0.7494 | 3/4 |
| Pluto | 248.400 | 0.6634 | 2/3 | Iapetus | 79.322 | 0.2682 | 2/7 |
| **Jupiter** | | | | **Uranus** | | | |
| Metis | 0.295 | -- | -- | Portia | 0.513 | -- | -- |
| Amalthea | 0.498 | 0.5917 | 3/5 | Puck | 0.762 | 0.6736 | 2/3 |
| Thebe | 0.675 | 0.7386 | 3/4 | Miranda | 1.413 | 0.5390 | 4/7 |
| Io | 1.769 | 0.3813 | 3/8 | Ariel | 2.520 | 0.5608 | 4/7 |
| Europa | 3.551 | 0.4982 | 1/2 | Umbriel | 4.144 | 0.6082 | 3/5 |
| Ganymede | 7.155 | 0.4964 | 1/2 | Titania | 8.706 | 0.4760 | 1/2 |
| Callisto | 16.689 | 0.4287 | 3/7 | Oberon | 13.463 | 0.6466 | 2/3 |
|  |  |  |  | **Neptune** | | | |
|  |  |  |  | Despina | 0.335 | -- | -- |
|  |  |  |  | Galatea | 0.429 | 0.7809 | 4/5 |
|  |  |  |  | Larissa | 0.555 | 0.7730 | 4/5 |
|  |  |  |  | Proteus | 1.122 | 0.4947 | 1/2 |
|  |  |  |  | Nereid | 360.130 | 0.0031 | << |

(*) Orbital periods in years for planetary system, in days for satellite systems;
<< corresponds to ratios smaller than 0.1181.

We see from Table 3 that for the Solar planetary system, the nine successive orbital period ratios are all close to fractions of Fibonacci numbers, with four times 2/5, twice 1/2, followed by 5/8, 1/3 and 2/3 once each. For the seven main moons of the Jupiter system, four of the six period ratios are close to fractions of Fibonacci numbers, i.e. 1/2 twice and 3/5 and 3/8 once each. For the ten main satellites of the Saturn system, four of the nine period ratios are close to fractions of Fibonacci numbers, i.e. 2/3 twice, 1/1 and 3/5 once each. For the seven main satellites of the Uranus system, four of the six period ratios are close to fractions of Fibonacci numbers, i.e. 2/3 twice, 3/5 and 1/2. For the five main satellites of the Neptune system (recall that the largest satellite Triton is retrograde), one ratio is close to a fraction of Fibonacci numbers, i.e. 1/2, while one ratio is much smaller than the set minimum of 1/8. In total of the five systems, one observes 22 orbital period ratios close to fractions of Fibonacci numbers, 11 ratios close to fractions involving one or no Fibonacci numbers and one ratio much smaller than the set minimum.

Figure 2 summarizes these findings.

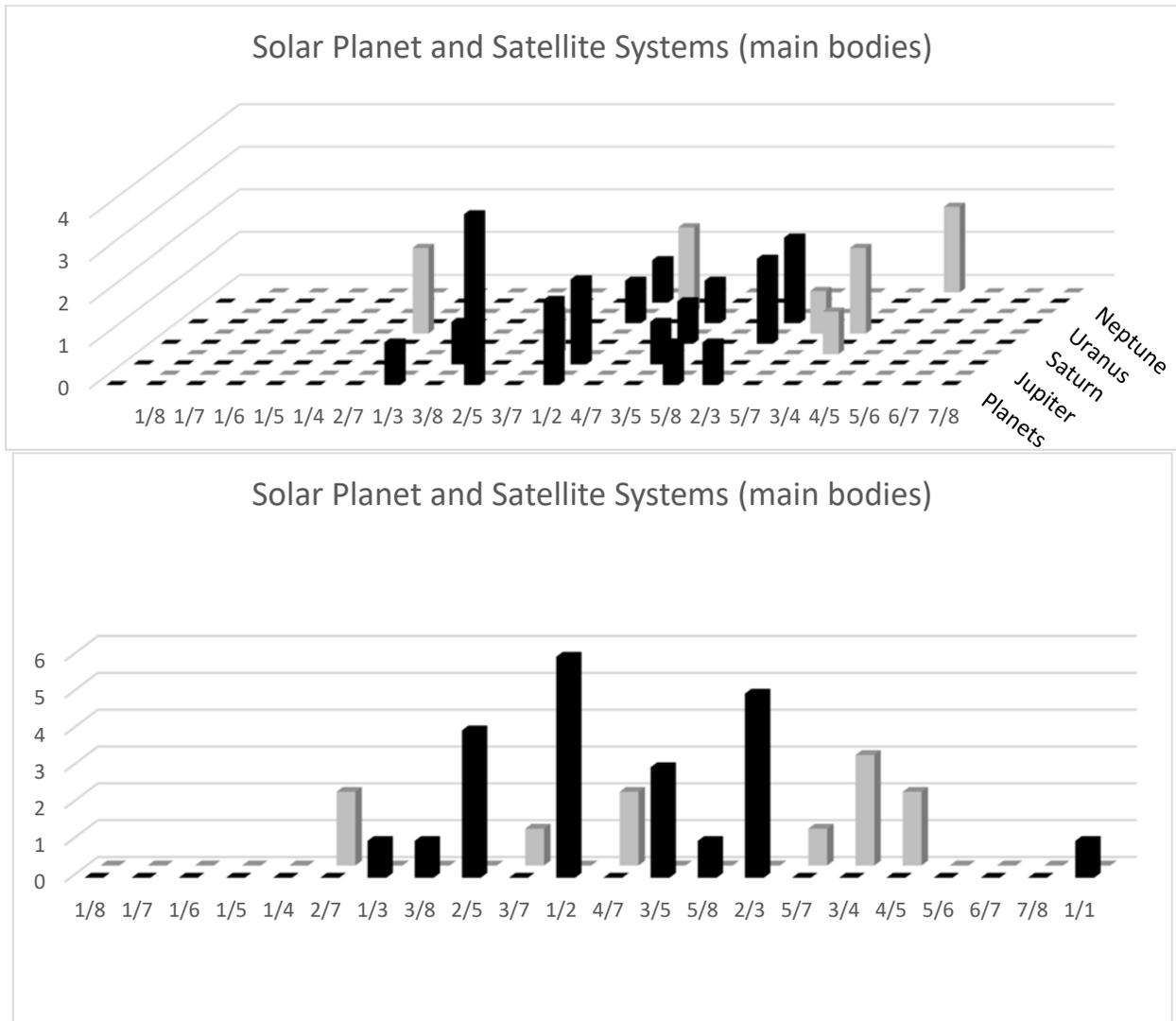

Figure 2: Histograms of numbers of period ratios for the Solar planetary and giant planet satellite systems close to irreducible fractions of two Fibonacci numbers (black) and of one or no Fibonacci numbers (grey).

3.2 <u>Planets and minor planets</u>

The catalogue of the Minor Planet Center (2017) listed in August 2017, 740 419 minor planets, including planetoids, asteroids, trojans, centaurs, Kuiper belt objects and other trans-Neptunian objects (TNOs). Only minor planets on prograde and elliptical orbits are considered; therefore 174 minor planets with inclination $i > 90°$ and/or eccentricity $e > 1$, are removed. Among the remaining 740 245 minor planets with $i < 90°$ and $e < 1$, 727 927 (98%) are asteroids with semi-major axis between 1.5 and 5.2 AU and therefore mainly gravitationally influenced by Jupiter and, to a lesser extent, by Mars. However, we consider also those minor planets whose orbit semi-major axis is less than 1.5 AU and greater than 5.2 AU as long as their period ratio with respectively Mars and Jupiter is comprised between 0.1181 and 1.

The ratios of periods $p_i$ of minor planets and of Jupiter and Mars have been compiled (Pletser, 2017a) such that the period ratios are always smaller than unity, i.e. $r_i = \frac{p_i}{p_{planet}}$ if $p_i < p_{planet}$ or $r_i = \frac{p_{planet}}{p_i}$ if $p_i > p_{planet}$, where $p_i$ is the orbital period of the $i^{th}$ minor planet and $p_{planet}$ refers to Jupiter or Mars orbital period. These ratios are then placed in boxes corresponding to the closest irreducible fraction of integers between 1 and 8 like above, except if the ratio is smaller than the inferior limit $r_{inf} = 0.1181$.

The following results are observed. There are 734 427 minor planets whose period ratios with Jupiter's range from 0.1181 to 1 and then to 0.1181, with semi-major axes between 1.25 and 21.60 AU. Similarly, there are 737 618 minor planets whose period ratios with Mars' range from 0.2167 to 1 and then to 0.1183, with semi-major axes between 0.55 and 6.32 AU. One obtains a total of 450 574 (61.35%) ratios close to fractions of two Fibonacci numbers and 283 853 (38.65%) ratios close to other fractions for ratios with Jupiter's period and respectively 438 191 (59.41%) and 299 427 (40.59%) for ratios with Mars' period, distributed as shown in Table 4 and Figure 3.

TABLE 4: NUMBER OF PERIOD RATIOS CLOSE TO FRACTIONS FOR MINOR PLANETS WITH JUPITER AND MARS, AND FOR TNOS AND NEPTUNE

| Closest fraction | Jupiter and minor planets | Mars and minor planets | Neptune and TNOs |
|---|---|---|---|
| 1/8 | 1167 | 21 | 6 |
| 1/7 | *1534* | *621* | *10* |
| 1/6 | *2008* | *6369* | *18* |
| 1/5 | 15004 | 130 | 26 |
| 1/4 | *16121* | *4722* | *42* |
| 2/7 | *146937* | *2662* | *37* |
| 1/3 | 150208 | 137009 | 45 |
| 3/8 | 142437 | 62611 | 34 |
| 2/5 | 51124 | 62920 | 73 |
| 3/7 | *115682* | *184790* | *57* |
| 1/2 | 79116 | 157145 | 257 |
| 4/7 | *1173* | *77608* | *488* |
| 3/5 | 113 | 8629 | 165 |
| 5/8 | 510 | 2200 | 149 |
| 2/3 | 4073 | 6103 | 224 |
| 5/7 | *115* | *12356* | *20* |
| 3/4 | *51* | *5313* | *26* |
| 4/5 | *32* | *1672* | *12* |
| 5/6 | *20* | *912* | *6* |
| 6/7 | *18* | *605* | *3* |
| 7/8 | *162* | *1797* | *7* |
| 1/1 | 6822 | 1423 | 11 |
| Sums | 450574 | 438191 | 990 |
|  | *283853* | *299427* | *726* |
| % of Total | 61.35 | 59.41 | 57.69 |
|  | *38.65* | *40.59* | *42.31* |

For each case, numbers of period ratios close to a fraction of Fibonacci numbers (left justified) and to other fractions (in italics, right justified)

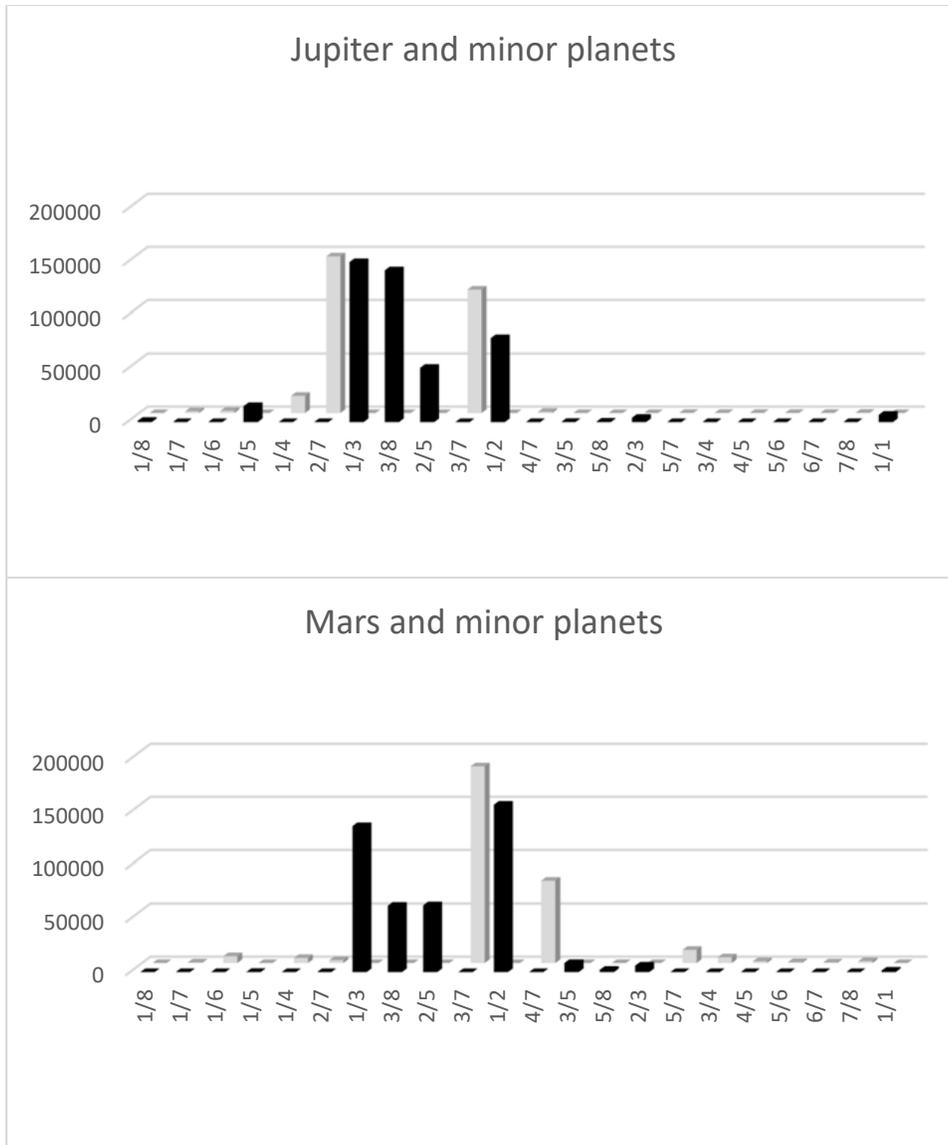

Figure 3: Histograms of numbers of period ratios of all minor planets with Jupiter and Mars close to irreducible fractions of two Fibonacci numbers (black) and of other fractions (grey).

It is seen that in both cases of Mars and Jupiter, the total numbers of period ratios close to irreducible fractions of Fibonacci numbers are always larger than those close to other fractions, in an approximate ratio 60% - 40%. If we look in more details for period ratios of Jupiter and minor planets, the highest value is for the fraction 1/3 (150 212) followed by 2/7 (146 940) and 3/8 (142 437), all close to each other. For Mars and minor planets, the highest value is for 3/7 (184 790) followed by 1/2 (157 146) and 1/3 (137 009).

3.3 Neptune and Trans-Neptunian Objects
The Minor Planet Center's list (2017) included in August 2017, 1772 Trans-Neptunian Objects (TNOs) with semi-major axis larger than Neptune's semi-major axis (30.11 AU). Among these, 9 are retrograde ($i > 90°$) and are removed, leaving 1763 TNOs, of which 1716 have their ratio of period with Neptune's period comprised between 1 and 0.1181 (see also Pletser, 2017a).

The period ratios of Neptune and TNOs show the same behaviour, i.e. a total number (990, 57.69%) of ratios close to irreducible fractions of Fibonacci numbers larger than those (726, 42.31%) close to other fractions as shown in Table 4 and Figure 4, despite the highest value being for 4/7 (488). This highest value is due to the fact that most of the TNOs are in radial position close to the 4/7 mean motion resonance with Neptune.

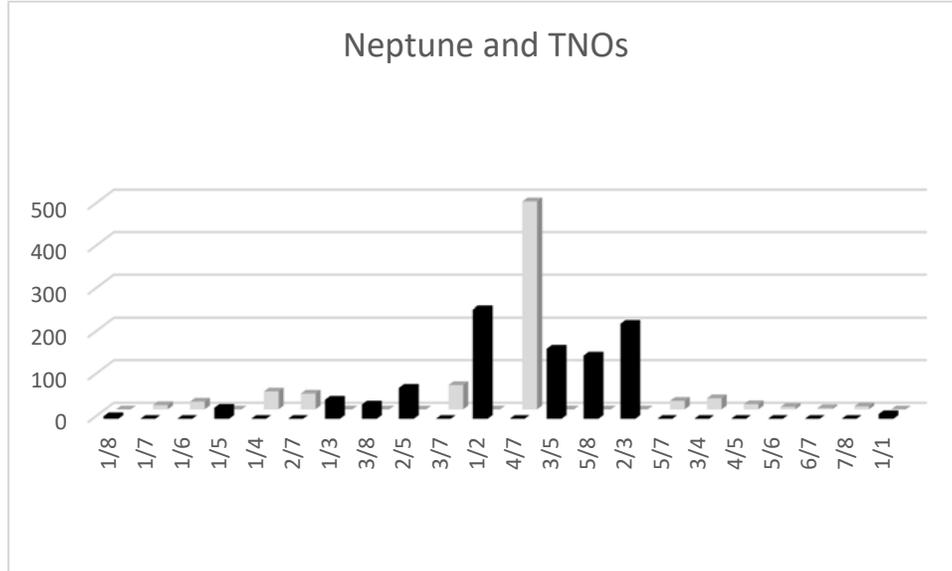

Figure 4: Histogram of the numbers of ratios of orbital periods of TNO's with Neptune close to irreducible fractions of two Fibonacci numbers (black) and other fractions (grey).
.
3.4 Exoplanetary systems

The European catalogue of exo-planets (Exoplanet TEAM, 2017) listed in August 2017, 3652 exoplanets, including 606 multiplanetary systems, i.e. 2 systems with 7 planets, 5 systems with 6 planets, 17 systems with 5 planets, 46 systems with 4 planets, 131 systems with 3 planets, and 405 systems with 2 planets.

Among those systems with two or more planets, one can test also our hypothesis of a preference for orbital period ratios close to irreducible fractions of Fibonacci numbers compared to other fractions. The ratios of the periods of successive exoplanets in their systems have been compiled (Pletser, 2017b) such that the period ratios are always smaller than unity.

Among these 910 pairs of consecutive exoplanets, one finds 791 pairs of consecutive planets with ratios between the inferior limit $r_{inf}$ = 0.1181 and 1, 115 pairs with ratios less than 0.1181 and 4 pairs of planets for which orbital elements are not yet determined. Table 5 shows the orbital periods of consecutive exoplanets, their ratio and the closest fraction for the 2 systems with 7 planets, the 5 systems with 6 planets, and the 17 systems with 5 planets. Note that the most recent value of the period for Planet Trappist-1 h is from (Luger et al., 2017).

TABLE 5: ORBITAL PERIODS, RATIOS AND CLOSEST FRACTIONS
FOR EXOPLANETARY SYSTEMS WITH 7, 6 AND 5 PLANETS

| | Orbital Period (days) | Ratio $r_i = \dfrac{p_{i-1}}{p_i}$ | Closest Fraction | | Orbital Period (days) | Ratio $r_i = \dfrac{p_{i-1}}{p_i}$ | Closest Fraction |
|---|---|---|---|---|---|---|---|
| **7 planets** | | | | **7 planets** | | | |
| Trappist-1 b | 1.511 | -- | -- | Kepler-90 b | 7.008 | -- | -- |
| Trappist-1 c | 2.422 | 0.6239 | 5/8 | Kepler-90 c | 8.719 | 0.8037 | 4/5 |
| Trappist-1 d | 4.050 | 0.5980 | 3/5 | Kepler-90 d | 59.737 | 0.1460 | 1/7 |
| Trappist-1 e | 6.100 | 0.6639 | 2/3 | Kepler-90 e | 91.939 | 0.6497 | 2/3 |
| Trappist-1 f | 9.207 | 0.6625 | 2/3 | Kepler-90 f | 124.914 | 0.7360 | 3/4 |
| Trappist-1 g | 12.353 | 0.7453 | 3/4 | Kepler-90 g | 210.607 | 0.5931 | 3/5 |
| Trappist-1 h | 18.767 | 0.6582 | 2/3 | Kepler-90 h | 331.601 | 0.6351 | 5/8 |
| **6 planets** | | | | **6 planets** | | | |
| GJ 667 C b | 7.200 | -- | -- | Kepler-11 b | 10.304 | -- | -- |
| GJ 667 C c | 28.140 | 0.2559 | 1/4 | Kepler-11 c | 13.025 | 0.7911 | 4/5 |
| GJ 667 C f | 39.026 | 0.7211 | 5/7 | Kepler-11 d | 22.687 | 0.5741 | 4/7 |
| GJ 667 C e | 62.240 | 0.6270 | 5/8 | Kepler-11 e | 31.996 | 0.7091 | 5/7 |
| GJ 667 C d | 91.610 | 0.6794 | 2/3 | Kepler-11 f | 46.689 | 0.6853 | 2/3 |
| GJ 667 C g | 256.200 | 0.3576 | 3/8 | Kepler-11 g | 118.378 | 0.3944 | 2/5 |
| HD 10180 c | 5.760 | -- | -- | Kepler-20 b | 3.696 | -- | -- |
| HD 10180 d | 16.358 | 0.3521 | 1/3 | Kepler-20 e | 6.099 | 0.6061 | 3/5 |
| HD 10180 e | 49.745 | 0.3288 | 1/3 | Kepler-20 c | 10.854 | 0.5619 | 4/7 |
| HD 10180 f | 122.760 | 0.4052 | 2/5 | Kepler-20 f | 19.578 | 0.5544 | 4/7 |
| HD 10180 g | 601.200 | 0.2042 | 1/5 | Kepler-20 g | 34.940 | 0.5603 | 4/7 |
| HD 10180 h | 2222.000 | 0.2706 | 2/7 | Kepler-20 d | 77.611 | 0.4502 | 3/7 |
| HD 40307 b | 4.312 | -- | -- | | | | |
| HD 40307 c | 9.621 | 0.4481 | 3/7 | | | | |
| HD 40307 d | 20.418 | 0.4712 | 1/2 | | | | |
| HD 40307 e | 34.620 | 0.5898 | 3/5 | | | | |
| HD 40307 f | 51.560 | 0.6715 | 2/3 | | | | |
| HD 40307 g | 197.800 | 0.2607 | 1/4 | | | | |
| **5 planets** | | | | **5 planets** | | | |
| 55 Cnc e | 0.737 | -- | -- | Kepler-292 b | 2.581 | -- | -- |
| 55 Cnc b | 14.653 | 0.0503 | << | Kepler-292 c | 3.715 | 0.6946 | 5/7 |
| 55 Cnc c | 44.373 | 0.3302 | 1/3 | Kepler-292 d | 7.056 | 0.5266 | 1/2 |
| 55 Cnc f | 260.910 | 0.1701 | 1/6 | Kepler-292 e | 11.979 | 0.5890 | 3/5 |
| 55 Cnc d | 4867.000 | 0.0536 | << | Kepler-292 f | 20.834 | 0.5750 | 4/7 |
| HD 219134 c | 6.764 | -- | -- | Kepler-296 b | 3.621 | -- | -- |
| HD 219134 d | 46.710 | 0.1448 | 1/7 | Kepler-296 c | 5.842 | 0.6199 | 5/8 |
| HD 219134 g | 94.200 | 0.4959 | 1/2 | Kepler-296 d | 19.850 | 0.2943 | 2/7 |
| HD 219134 e | 1190.000 | 0.0792 | << | Kepler-296 e | 34.142 | 0.5814 | 4/7 |
| HD 219134 h | 2198.000 | 0.5414 | 4/7 | Kepler-296 f | 63.336 | 0.5391 | 4/7 |
| HIP 41378 b | 15.571 | -- | -- | Kepler-33 b | 5.668 | -- | -- |
| HIP 41378 c | 31.698 | 0.4912 | 1/2 | Kepler-33 c | 13.176 | 0.4302 | 3/7 |
| HIP 41378 e | 131.000 | 0.2420 | 1/4 | Kepler-33 d | 21.776 | 0.6051 | 3/5 |
| HIP 41378 d | 157.000 | 0.8344 | 5/6 | Kepler-33 e | 31.784 | 0.6851 | 2/3 |
| HIP 41378 f | 324.000 | 0.4846 | 1/2 | Kepler-33 f | 41.029 | 0.7747 | 3/4 |
| Kepler-102 b | 5.287 | -- | -- | Kepler-444 b | 3.600 | -- | -- |
| Kepler-102 c | 7.071 | 0.7477 | 3/4 | Kepler-444 c | 4.546 | 0.7919 | 4/5 |
| Kepler-102 d | 10.312 | 0.6858 | 2/3 | Kepler-444 d | 6.189 | 0.7345 | 3/4 |
| Kepler-102 e | 16.146 | 0.6387 | 5/8 | Kepler-444 e | 7.743 | 0.7993 | 4/5 |
| Kepler-102 f | 27.454 | 0.5881 | 3/5 | Kepler-444 f | 9.740 | 0.7950 | 4/5 |
| Kepler-150 b | 3.428 | -- | -- | Kepler-55 d | 2.211 | -- | -- |
| Kepler-150 c | 7.382 | 0.4644 | 1/2 | Kepler-55 e | 4.618 | 0.4788 | 1/2 |
| Kepler-150 d | 12.561 | 0.5877 | 3/5 | Kepler-55 f | 10.199 | 0.4528 | 3/7 |
| Kepler-150 e | 30.827 | 0.4075 | 2/5 | Kepler-55 b | 27.948 | 0.3649 | 3/8 |
| Kepler-150 f | 637.209 | 0.0484 | << | Kepler-55 c | 42.152 | 0.6630 | 2/3 |
| Kepler-154 e | 3.933 | -- | -- | Kepler-62 b | 5.715 | -- | -- |
| Kepler-154 f | 9.919 | 0.3965 | 2/5 | Kepler-62 c | 12.442 | 0.4593 | 3/7 |
| Kepler-154 d | 20.550 | 0.4827 | 1/2 | Kepler-62 d | 18.164 | 0.6850 | 2/3 |
| Kepler-154 b | 33.041 | 0.6220 | 5/8 | Kepler-62 e | 122.387 | 0.1484 | 1/7 |
| Kepler-154 c | 62.303 | 0.5303 | 1/2 | Kepler-62 f | 267.291 | 0.4579 | 3/7 |
| Kepler-169 b | 3.251 | -- | -- | Kepler-80 f | 0.990 | -- | -- |
| Kepler-169 c | 6.195 | 0.5247 | 1/2 | Kepler-80 d | 3.070 | 0.3225 | 1/3 |
| Kepler-169 d | 8.348 | 0.7421 | 3/4 | Kepler-80 e | 4.640 | 0.6616 | 2/3 |
| Kepler-169 e | 13.767 | 0.6064 | 3/5 | Kepler-80 b | 7.050 | 0.6582 | 2/3 |
| Kepler-169 f | 87.090 | 0.1581 | 1/6 | Kepler-80 c | 9.520 | 0.7405 | 3/4 |
| Kepler-186 b | 3.887 | -- | -- | Kepler-84 d | 4.225 | -- | -- |
| Kepler-186 c | 7.267 | 0.5348 | 1/2 | Kepler-84 b | 8.726 | 0.4841 | 1/2 |
| Kepler-186 d | 13.343 | 0.5446 | 4/7 | Kepler-84 c | 12.883 | 0.6773 | 2/3 |
| Kepler-186 e | 22.408 | 0.5954 | 3/5 | Kepler-84 e | 27.434 | 0.4696 | 1/2 |
| Kepler-186 f | 129.946 | 0.1724 | 1/6 | Kepler-84 f | 44.552 | 0.6158 | 5/8 |
| Kepler-238 b | 2.091 | -- | -- | | | | |
| Kepler-238 c | 6.156 | 0.3397 | 1/3 | | | | |
| Kepler-238 d | 13.234 | 0.4651 | 1/2 | | | | |
| Kepler-238 e | 23.654 | 0.5595 | 4/7 | | | | |
| Kepler-238 f | 50.447 | 0.4689 | 1/2 | | | | |

<< corresponds to ratios smaller than 0.1181.

It is remarkable that for the Trappist-1 system, five of the six ratios are close to Fibonacci fractions (Gillon et al., 2017; Luger et al., 2017; Pletser and Basano, 2017). For the Kepler-90 system, three of the six ratios are close to Fibonacci fractions, yielding a total of 8 out of 12 ratios (66.67%) close to Fibonacci fractions. For the five GJ 667 C, HD 10180, HD 40307, Kepler-11 and Kepler-20 exoplanetary systems with 6 planets, respectively 3, 4, 3, 2 and 1 out of 5 ratios are close to Fibonacci fractions, yielding a total of 13 out of 25 ratios (52%) close to Fibonacci fractions. For the 17 exoplanetary systems with 5 planets, a total of 37 out of 64 ratios (57.81%) are close to Fibonacci fractions, disregarding the four cases of ratios smaller than 0.1181 and despite one system (Kepler-444) having no ratio close to a Fibonacci fraction at all and four other systems (55 Cnc, HD 219134, Kepler-296, Kepler-62) having only one ratio close to a Fibonacci fraction. For the 46 exoplanetary systems with 4 planets, a total of 72 out of 129 ratios (55.81%) are close to Fibonacci fractions, disregarding 9 cases of ratios smaller than 0.1181. For the 131 exoplanetary systems with 3 planets, a total of 155 out of 242 ratios (64.05%) are close to Fibonacci fractions, disregarding 20 cases of ratios smaller than 0.1181. Finally, for the 401 systems with two planets, a total of 188 out of 319 ratios (58.93%) are close to Fibonacci fractions, disregarding 82 cases of ratios smaller than 0.1181. The distribution of period ratios close to Fibonacci and other fractions is given in Table 6 and Figure 5 for all 791 exoplanetary systems of 2 to 7 planets with orbital period ratios between 0.1181 and 1.

TABLE 6: NUMBER OF PERIOD RATIOS CLOSE TO
FRACTIONS IN ALL EXOPLANETARY SYSTEMS

| Closest fraction | Exoplanetary Systems | |
|---|---|---|
| 1/8 | 7 | |
| *1/7* | | *18* |
| *1/6* | | *20* |
| 1/5 | 34 | |
| *1/4* | | *39* |
| *2/7* | | *35* |
| 1/3 | 52 | |
| 3/8 | 50 | |
| 2/5 | 45 | |
| *3/7* | | *84* |
| 1/2 | 135 | |
| *4/7* | | *61* |
| 3/5 | 41 | |
| 5/8 | 41 | |
| 2/3 | 66 | |
| *5/7* | | *27* |
| *3/4* | | *16* |
| *4/5* | | *12* |
| *5/6* | | *3* |
| *6/7* | | *2* |
| *7/8* | | *1* |
| 1/1 | 2 | |
| Sums | 473 | |
| | | *318* |
| % of Total | 59.80 | |
| | | *40.20* |

Numbers of period ratios close
to a fraction of Fibonacci numbers (left justified)
and to other fractions (in italics, right justified)

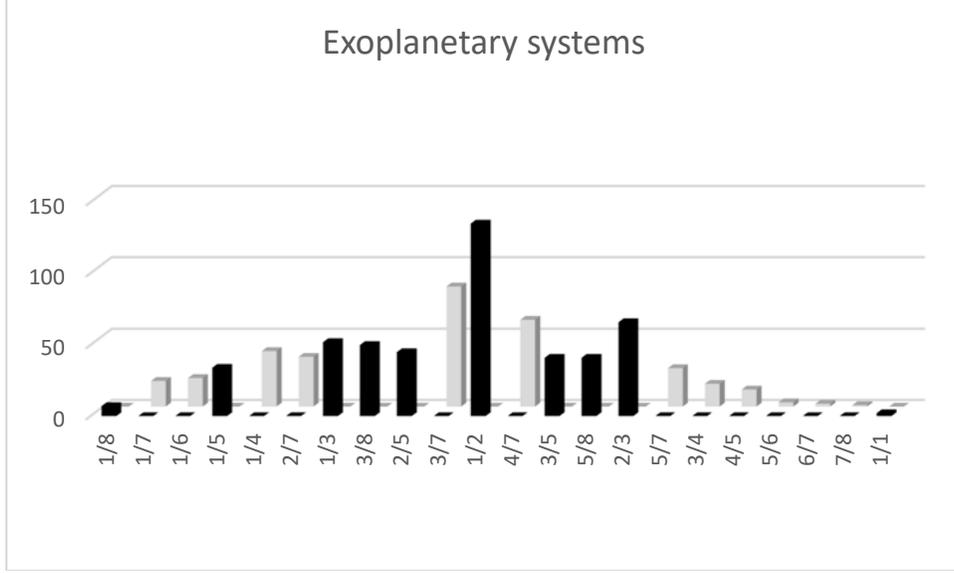

Figure 5: Histogram of the numbers of ratios of orbital periods of all adjacent planets in all exoplanetary systems, close to irreducible fractions of two Fibonacci numbers (black) and other fractions (grey).

It is seen that the highest peak is observed for 1/2 with 135 cases, followed by 3/7 with 84 cases and 2/3 with 66 cases, and in total, there are 473 out of 791 ratios (59.80%) close to Fibonacci fractions, disregarding 115 cases of ratios smaller than 0.1181.

### 4. From all minor planets to "regular" ones

For the 740 245 minor planets on prograde and elliptic orbits considered in section 3.2, considering all minor planets in range of Jupiter and Mars at the same level of interest for our discussion, as done in Section 3.2, is somehow misleading. This means that the distribution of their period or, from Kepler's third law, of their semi-major axis, is considered only without due consideration for other orbital elements, mainly inclination and eccentricity. And indeed, if we are looking at stabilizing gravitational interactions between Jupiter and, to a lesser extent Mars, and minor planets' orbits, let us consider only those minor planets that have their orbits nearly coplanar with the ecliptic and nearly circularized. For argument's sake, next to the first case with all minor planets with $i < 90°$ and $e < 1$, we consider the nine following cases of minor planets with inclinations and eccentricities $(i, e)$ less than respectively (10°, 0.2), (10°, 0.1), (5°, 0.1), (5°, 0.05), (3°, 0.05), (3°, 0.03), (2°, 0.03), (2°, 0.02) and (2°, 0.01), whose orbits are more "regular", i.e. less inclined and more circularized, when passing from one case to the next. As comparison, recall that Mercury's orbit has an eccentricity of 0.206 and is inclined by 7° on the ecliptic. Filtering successively the list of minor planets, for each of above pairs of values of $(i, e)$, one obtains $N$ "regular" minor planets, among which $N_J$ and $N_M$ are in range of respectively Jupiter and Mars, i.e. with period ratios ranging from 0.1181 to 1 to 0.1181. It yields a total of $N_{fJ}$ and $N_{fM}$ ratios close to fractions of two Fibonacci numbers and $N_{oJ}$ and $N_{oM}$ ratios close to other fractions (see Table 7). For example, for the second case $(i, e) < (10°, 0.2)$, one obtains $N = 371\ 498$ more "regular" minor planets with $i < 10°$ and $e < 0.2$, among which $N_J = 369\ 982$ are in range of Jupiter, i.e. with period ratios ranging from 0.1186 to 1 to 0.1182, corresponding to semi-major axis ranging between 1.25 and 21.59 AU, and $N_M = 370\ 546$ "regular" minor planets in range of Mars, i.e. with period ratios ranging from 0.4141 to 1 to 0.1395, corresponding to semi-major axis ranging between 0.85 and 5.66 AU. It yields a total of $N_{fJ}$ = 209 222 (56.55%) ratios close to fractions of two Fibonacci numbers and $N_{oJ}$ = 160 752 (43.45%) ratios close to other fractions for ratios with Jupiter's period and respectively $N_{fM}$ = 234 175 (63.20%) and $N_{oM}$ = 136 371 (36.80%) for ratios with Mars' period, distributed as shown in Table 7.

TABLE 7: NUMBER OF PERIOD RATIOS CLOSE TO FRACTIONS
FOR "REGULAR" MINOR PLANETS AND JUPITER AND MARS

| Minor Planets | | Total | Jupiter | | | Mars | | |
|---|---|---|---|---|---|---|---|---|
| $i <$ | $e <$ | $N$ | $N_J$ | $N_{fJ}$ | $N_{oJ}$ | $N_M$ | $N_{fM}$ | $N_{oM}$ |
| 90° | 1 | 740 245 | 734 427 | 450 574 (61.35%) | 283 853 (38.65%) | 737 618 | 438 191 (59.41%) | 299 427 (40.59%) |
| 10° | 0.2 | 371 498 | 369 974 | 209 222 (56.55%) | 160 752 (43.45%) | 370 546 | 234 175 (63.20%) | 136 371 (36.80%) |
| 10° | 0.1 | 132 879 | 132 081 | 81 059 (61.37%) | 51 022 (38.63%) | 132 229 | 85 910 (64.97%) | 46 319 (35.03%) |
| 5° | 0.1 | 60 253 | 59 644 | 39 531 (66.28%) | 20 113 (33.72%) | 59 722 | 36 975 (61.91%) | 2 2747 (38.09%) |
| 5° | 0.05 | 16 916 | 16 600 | 12 001 (72.30%) | 4 599 (27.70%) | 16 620 | 10 717 (64.48%) | 5 903 (35.52%) |
| 3° | 0.05 | 8 238 | 8 008 | 5 390 (67.31%) | 2 618 (32.69%) | 8 020 | 5 441 (67.84%) | 2 579 (32.16%) |
| 3° | 0.03 | 3 347 | 3 211 | 2 221 (69.17%) | 990 (30.83%) | 3 215 | 2 245 (69.83%) | 970 (30.17%) |
| 2° | 0.03 | 1 662 | 1 587 | 1 078 (67.93%) | 509 (32.07%) | 1 589 | 1 147 (72.18%) | 442 (27.82%) |
| 2° | 0.02 | 820 | 763 | 522 (68.41%) | 241 (31.59%) | 763 | 579 (75.88%) | 184 (24.12%) |
| 2° | 0.01 | 222 | 182 | 128 (70.33%) | 54 (29.67%) | 182 | 129 (70.88%) | 53 (29.12%) |

The distribution of number of ratios close to fractions for the ten cases of all minor planets to "regular" ones and Jupiter is given in Table 8. Figure 6 shows the distribution of the relative number of ratios, i.e. the number of ratios divided by the total number of ratios. It is seen that moving from the set of all minor planets to more "regular" ones, the relative number of orbital period ratios close to the non-Fibonacci fractions 2/7 and 3/7 decrease while those close to the Fibonacci fractions 3/8 and 2/5 have a tendency to increase (see also Figure 7).

In total, the number of orbital period ratios close to all Fibonacci fractions increase from approximately 61% to 70%, passing through a dip at 56.5% for the second case of $(i, e) <$ $(10°, 0.2)$.

TABLE 8: NUMBER OF PERIOD RATIOS CLOSE TO FRACTIONS
FOR ALL MINOR PLANETS TO MORE "REGULAR" ONES AND JUPITER

| Closest fraction | $i < 90°$ $e < 1$ | $i < 10°$ $e < 0.2$ | $i < 10°$ $e < 0.1$ | $i < 5°$ $e < 0.1$ | $i < 5°$ $e < 0.05$ | $i < 3°$ $e < 0.05$ | $i < 3°$ $e < 0.03$ | $i < 2°$ $e < 0.03$ | $i < 2°$ $e < 0.02$ | $i < 2°$ $e < 0.01$ |
|---|---|---|---|---|---|---|---|---|---|---|
| 1/8 | 1167 | 78 | 11 | 3 | 1 | 0 | 0 | 0 | 0 | 0 |
| *1/7* | *1534* | *36* | *9* | *3* | *0* | *0* | *0* | *0* | *0* | *0* |
| *1/6* | *2008* | *54* | *32* | *10* | *5* | *0* | *0* | *0* | *0* | *0* |
| 1/5 | 15004 | 108 | 44 | 15 | 8 | 2 | 1 | 1 | 1 | 1 |
| *1/4* | *16121* | *3551* | *1063* | *847* | *238* | *127* | *53* | *24* | *14* | *0* |
| *2/7* | *146937* | *103370* | *23775* | *9179* | *1474* | *613* | *195* | *97* | *41* | *0* |
| 1/3 | 150208 | 82494 | 17645 | 7848 | 1758 | 825 | 278 | 145 | 66 | 17 |
| 3/8 | 142437 | 60978 | 28521 | 15003 | 4712 | 1971 | 798 | 344 | 156 | 38 |
| 2/5 | 51124 | 32344 | 20573 | 13194 | 4610 | 2169 | 1001 | 517 | 274 | 65 |
| *3/7* | *115682* | *53209* | *25777* | *9999* | *2857* | *1869* | *732* | *385* | *184* | *0* |
| 1/2 | 79116 | 28599 | 11508 | 2532 | 572 | 290 | 100 | 46 | 17 | 4 |
| *4/7* | *1173* | *501* | *356* | *73* | *24* | *8* | *5* | *3* | *2* | *0* |
| 3/5 | 113 | 17 | 13 | 11 | 7 | 5 | 1 | 0 | 0 | 0 |
| 5/8 | 510 | 81 | 30 | 0 | 11 | 4 | 3 | 2 | 1 | 1 |
| 2/3 | 4073 | 1462 | 219 | 129 | 25 | 13 | 4 | 2 | 0 | 0 |
| *5/7* | *115* | *0* | *0* | *27* | *0* | *0* | *0* | *0* | *0* | *0* |
| *3/4* | *51* | *4* | *2* | *1* | *1* | *1* | *0* | *0* | *0* | *0* |
| *4/5* | *32* | *0* | *0* | *0* | *0* | *0* | *0* | *0* | *0* | *0* |
| *5/6* | *20* | *0* | *0* | *0* | *0* | *0* | *0* | *0* | *0* | *0* |
| *6/7* | *18* | *3* | *0* | *0* | *0* | *0* | *0* | *0* | *0* | *0* |
| *7/8* | *162* | *24* | *10* | *1* | *0* | *0* | *0* | *0* | *0* | *0* |
| 1/1 | 6822 | 3061 | 2493 | 769 | 297 | 111 | 40 | 21 | 7 | 2 |
| Sums | 450574 | 209222 | 81059 | 39531 | 12001 | 5390 | 2221 | 1078 | 522 | 128 |
|  | *283853* | *160752* | *51022* | *20113* | *4599* | *2618* | *990* | *509* | *241* | *54* |
| % of Total | 61.35 | 56.55 | 61.37 | 66.28 | 72.30 | 67.31 | 69.17 | 67.93 | 68.41 | 70.33 |
|  | *38.65* | *43.45* | *38.63* | *33.72* | *27.70* | *32.69* | *30.83* | *32.07* | *31.59* | *29.67* |

For each case, numbers of period ratios close to a fraction of Fibonacci numbers (left justified) and to other fractions (in italics, right justified)

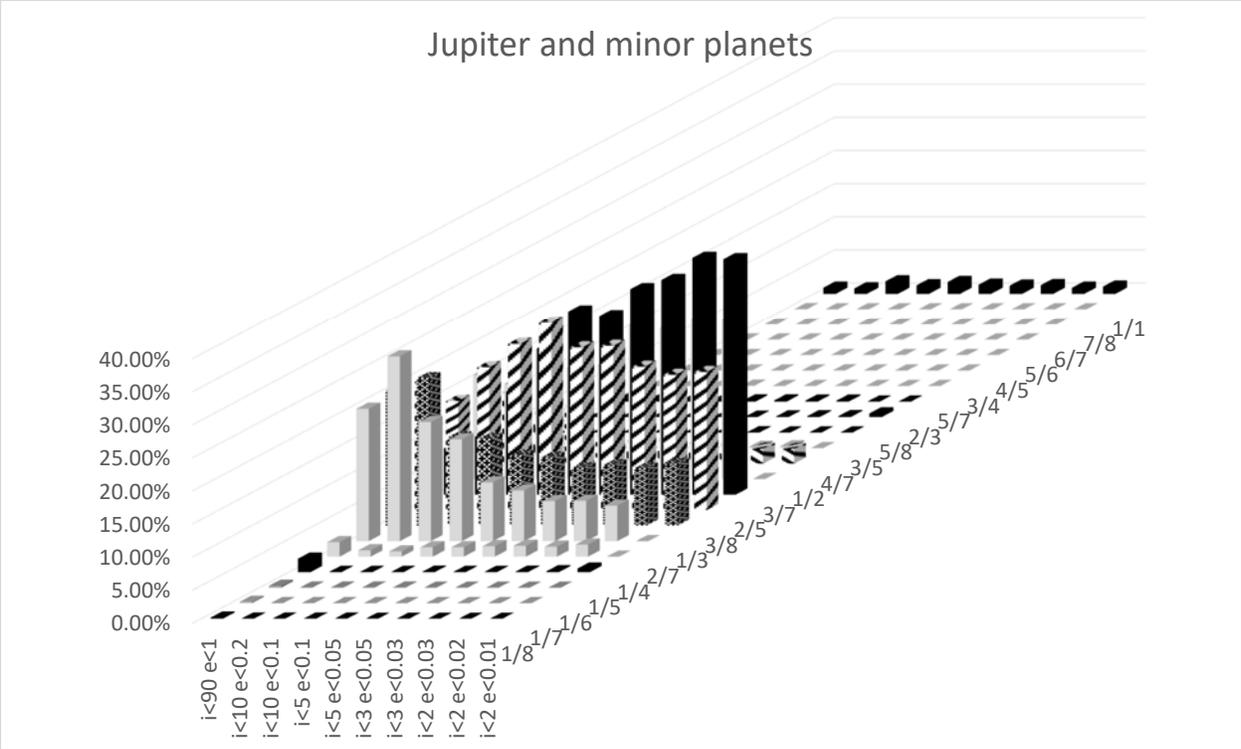

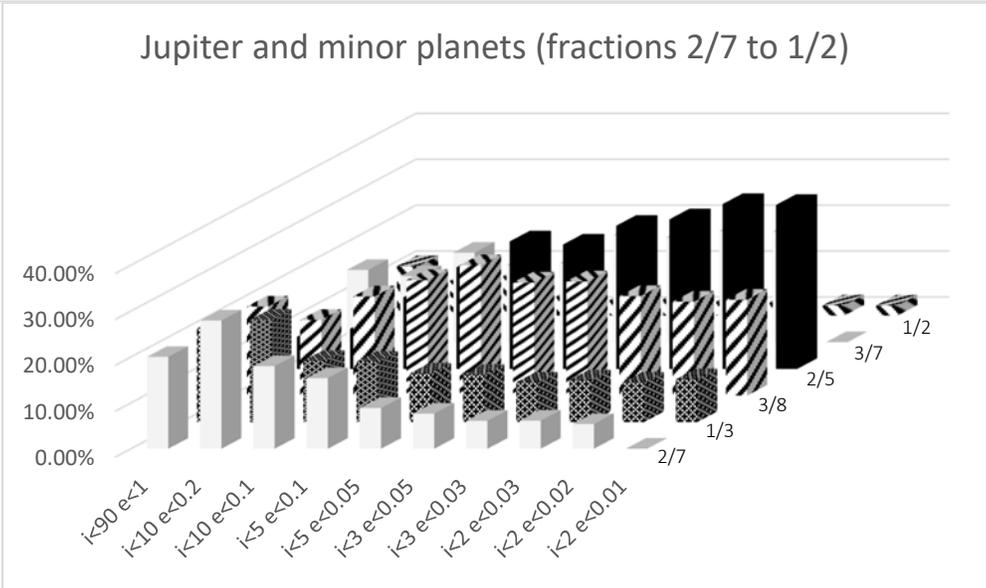

Figure 6: Histograms of the percentage of the number of ratios of orbital periods of minor planets with Jupiter, from all minor planets ($i < 90°, e < 1$, left) to "regular" minor planets ($i < 2°, e < 0.01$, right) for all fractions (top panel) and for fractions 2/7 to 1/2 (bottom panel).

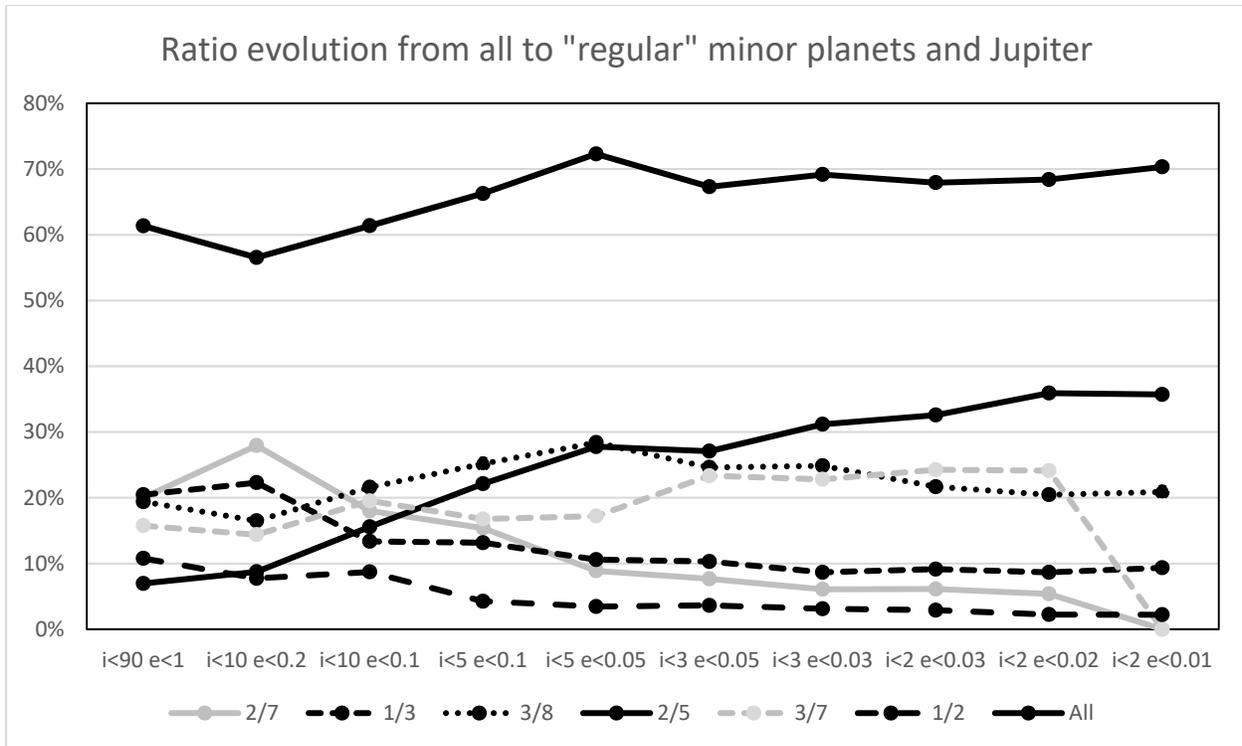

Figure 7: Evolution of relative numbers of period ratios of minor planets and Jupiter from all minor planets to "regular" ones for fractions 2/7 to 1/2 and for all period ratios close to Fibonacci fractions (black) and other fractions (grey).

Similarly, the distribution of number of ratios close to fractions for the ten cases of all minor planets to "regular" ones and Mars is given in Table 9 and shown in Figure 8. It is seen that moving from the set of all minor planets to more "regular" ones, the relative number of orbital ratios close to the non-Fibonacci fractions 4/7 and 3/7 tend respectively to decrease and to stay the same, while those close to the Fibonacci fractions 3/8 and 2/5 have a tendency to increase and those close to 1/2 tend to decrease.

In total, the number of orbital period ratios close to all Fibonacci fractions increase from approximately 59% to 71%, with a maximum of 75.9% for the case before last of $(i, e) < (2°, 0.02)$.

TABLE 9: NUMBER OF PERIOD RATIOS CLOSE TO FRACTIONS
FOR ALL MINOR PLANETS TO MORE "REGULAR" ONES AND MARS

| Closest fraction | $i < 90°$ $e < 1$ | $i < 10°$ $e < 0.2$ | $i < 10°$ $e < 0.1$ | $i < 5°$ $e < 0.1$ | $i < 5°$ $e < 0.05$ | $i < 3°$ $e < 0.05$ | $i < 3°$ $e < 0.03$ | $i < 2°$ $e < 0.03$ | $i < 2°$ $e < 0.02$ | $i < 2°$ $e < 0.01$ |
|---|---|---|---|---|---|---|---|---|---|---|
| 1/8 | 21 | 0 | 0 | 0 | 0 | 0 | 0 | 0 | 0 | 0 |
| *1/7* | *621* | *318* | *285* | *110* | *42* | *22* | *2* | *1* | *1* | *1* |
| *1/6* | *6369* | *2767* | *2218* | *660* | *255* | *89* | *38* | *20* | *6* | *1* |
| 1/5 | 130 | 8 | 2 | 1 | 1 | 1 | 0 | 0 | 0 | 0 |
| *1/4* | *4722* | *1557* | *255* | *160* | *38* | *18* | *7* | *4* | *1* | *1* |
| *2/7* | *2662* | *1040* | *702* | *143* | *53* | *24* | *9* | *4* | *2* | *1* |
| 1/3 | 137009 | 52181 | 20786 | 4934 | 1150 | 619 | 224 | 109 | 43 | 11 |
| 3/8 | 62611 | 32979 | 19160 | 9784 | 3103 | 2123 | 910 | 457 | 226 | 62 |
| 2/5 | 62920 | 39017 | 23377 | 13870 | 4735 | 1914 | 842 | 456 | 248 | 46 |
| *3/7* | *184790* | *73539* | *28576* | *15528* | *4552* | *2038* | *795* | *348* | *151* | *42* |
| 1/2 | 157145 | 103647 | 20608 | 6901 | 1355 | 598 | 197 | 92 | 41 | 4 |
| *4/7* | *77608* | *56825* | *14195* | *6116* | *949* | *387* | *118* | *64* | *22* | *6* |
| 3/5 | 8629 | 5487 | 1811 | 1403 | 354 | 176 | 68 | 31 | 19 | 4 |
| 5/8 | 2200 | 502 | 93 | 54 | 13 | 7 | 3 | 2 | 2 | 2 |
| 2/3 | 6103 | 320 | 55 | 23 | 5 | 3 | 1 | 0 | 0 | 0 |
| *5/7* | *12356* | *118* | *25* | *10* | *3* | *0* | *0* | *0* | *0* | *0* |
| *3/4* | *5313* | *70* | *15* | *5* | *4* | *1* | *1* | *1* | *1* | *1* |
| *4/5* | *1672* | *39* | *10* | *2* | *1* | *0* | *0* | *0* | *0* | *0* |
| *5/6* | *912* | *34* | *11* | *5* | *2* | *0* | *0* | *0* | *0* | *0* |
| *6/7* | *605* | *18* | *5* | *0* | *0* | *0* | *0* | *0* | *0* | *0* |
| *7/8* | *1797* | *46* | *22* | *8* | *4* | *0* | *0* | *0* | *0* | *0* |
| 1/1 | 1423 | 34 | 18 | 5 | 1 | 0 | 0 | 0 | 0 | 0 |
| Sums | 438191 | 234175 | 132229 | 36975 | 10717 | 5441 | 2245 | 1147 | 579 | 129 |
|  | *299427* | *136371* | *46319* | *22747* | *5903* | *2579* | *970* | *442* | *184* | *53* |
| % of Total | 59.41 | 63.20 | 74.06 | 61.91 | 64.48 | 67.84 | 69.83 | 72.18 | 75.88 | 70.88 |
|  | *40.59* | *36.80* | *25.94* | *38.09* | *35.52* | *32.16* | *30.17* | *27.82* | *24.12* | *29.12* |

For each case, numbers of period ratios close to a fraction of Fibonacci numbers (left justified) and to other fractions (in italics, right justified)

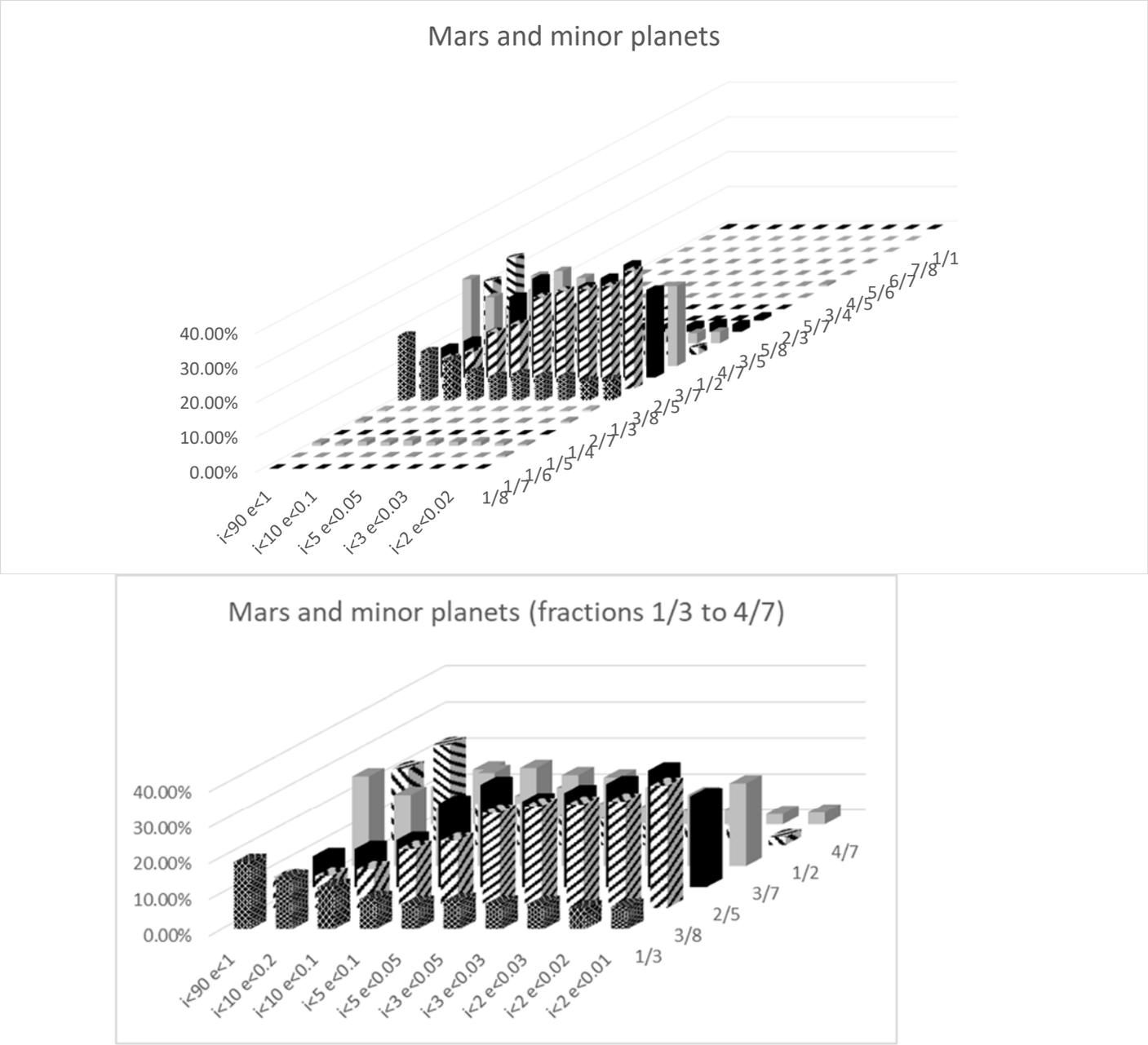

Figure 8: Histograms of the percentage of the number of ratios of orbital periods of minor planets with Mars, from all minor planets ($i < 90°, e < 1$, left) to "regular" minor planets ($i < 2°, e < 0.01$, right) for all fractions (top panel) and for fractions 1/3 to 4/7 (bottom panel).

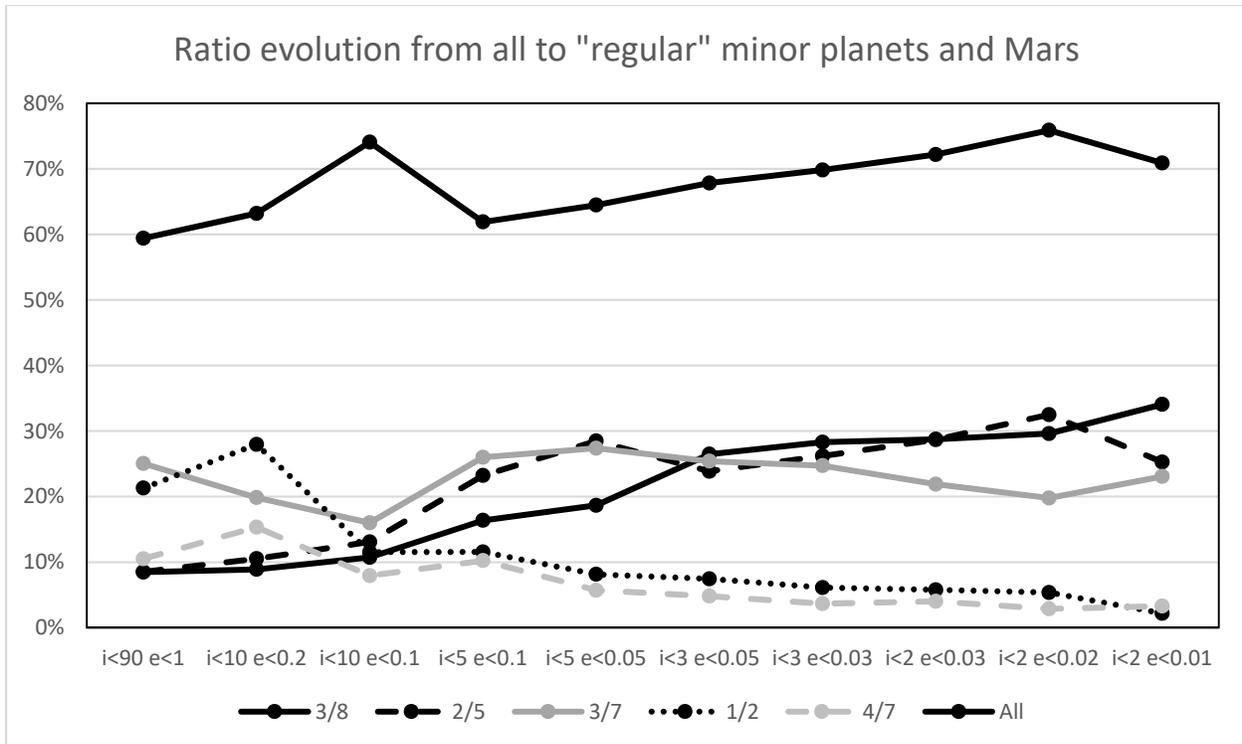

Figure 9: Evolution of relative numbers of period ratios of minor planets and Mars from all to "regular" minor planets for fractions 3/8 to 4/7 and for all period ratios close to Fibonacci fractions (black) and other fractions (grey).

## 5. Discussion
5.1 Solar and exoplanetary systems

It is seen that, for the solar planetary and satellite systems and exoplanetary systems in total, the number of period ratios close to fractions formed with two Fibonacci integers exceeds significantly the number of period ratios close to other fractions in a ratio of approximately 60% to 40 %, although there are individual cases in giant planet satellite systems and in some exoplanetary systems where the opposite occurs.

This is in contrast with what was simply calculated in Section 2, i.e. only a 45.4 % of chance of Fibonacci fractions appearing if all 22 irreducible fractions formed with integers from 1 to 8 would be uniformly distributed (which is of course not the case), or less, 40.1 % of chance, if the width of the associated bins is considered (see Table 2). So, we can easily conclude that orbital periods of successive secondaries (either planets or satellites) have a tendency to be closer to fractions of Fibonacci integers.

In particular, although none of the solar planets are in exact resonance, they all are close to mean motion resonance ratios given by irreducible fractions of Fibonacci numbers. For the Venus-Earth case, Bazsó et al. (2010) discusses the three possible quasi-resonances 3/5, 5/8 and 8/13, which are all Fibonacci ratios. The minor planets Pluto and the Plutinos are in a well-known mean motion 2/3 resonance with Neptune (see e.g. Peale, 1976; Murray and Dermott, 1999).

Among satellite systems of giant planets, there are additional resonances between non-successive satellites in the Saturnian system, namely the 1/2 resonances between Mimas and Tethys and between Enceladus and Dione (although the resonance between Mimas and Tethys is really a 2/4 resonance (see e.g. Goldreich, 1965; Roy and Ovenden, 1954), the ratio of their periods is still 0.942/1.888 = 0.499). In addition, there are also resonances between large and small satellites and between small satellites and rings that are not counted here. One should also add many cases of co-orbital small secondaries with larger planets or satellites in 1/1 resonances, like e.g. Trojan asteroids and Jupiter, and small Saturnian satellites co-orbital with Dione and Tethys (Murray and Dermott, 1999). In the Uranian system, there also some near resonances between small satellites and rings, like between Rosalind and Cordelia close to 3/5 (Murray and Thompson, 1990) and Cordelia and Ophelia with the edge of the epsilon ring by a 24/25 and 13/14 (Goldreich and Porco, 1987) or, according to our approach of using only integers between 1 and 8, both being close to 1/1.

For the asteroids, although there are concentrations of asteroids at 2/3 and 1/1 resonances with Jupiter, Kirkwood (1888) found gaps at mean motion resonances 1/4, 2/7, 1/3, 3/8, 2/5, 3/7, 1/2, 3/5, where most of asteroids have been ejected by repeated perturbations by Jupiter (Wisdom, 1982, 1983, 1985; Minton and Malhotra, 2009). So, it might be surprising to still find minor planets whose orbital period ratio with Jupiter is associated to these values. But recall that we put orbital period ratios in bins of a certain width around small integer fractions. So, for example, Ceres is not in an exact 2/5 resonance with Jupiter, although their period ratio is in the bin associated to 2/5.

Regarding exoplanets, it is striking to see that most of the multiplanetary exo-systems have so many pairs of planets with orbital period ratios close to Fibonacci fraction, e.g. five out of six ratios in the seven planet Trappist-1 system are close to Fibonacci fractions, while for other exo-systems, e.g. the five planet Kepler-444 system, none of the successive planet orbital period ratios are close to Fibonacci fractions.

In a separate study, Aschwanden and Scholkmann (2017) came to the conclusion that 73% of 932 exoplanets pairs have period ratios close to (2:1), (3:2), and (5:3), i.e. the second to the fifth Fibonacci numbers. Veras and Ford (2012) found that Kepler multiplanet systems preferentially cluster around commensurabilities $(j_1:|j_2|)$ (with integers $j_1 > |j_2|$) where $|j_2|$ is low, and rarely do so when $|j_2|$ is high. In particular, the number of Kepler systems near the (2:1), (3:2), (3:1) and (5:3) commensurabilities is higher than what would be expected from a random distribution of Kepler planet candidate semimajor axes. This analysis shows that a high fraction of exoplanetary systems may be near resonance but not actually in resonance.

5.2 From all to "regular" minor planets

Concerning asteroids, the analysis made in Section 4 with sets from all minor planets to more "regular" ones shows that, when moving from a heteroclite set of secondaries with large ranges of orbital eccentricities and inclinations to smaller but more "regular" sets of secondaries with smaller eccentricities and inclinations, the relative numbers of period ratios of minor planets with Jupiter and Mars are preferentially closer to irreducible fractions of Fibonacci integers. This is most likely a sign that orbital quasi-commensurabilities with period ratios close to Fibonacci fractions are associated with more regular orbits, less inclined and more circular, and/or of a natural evolution toward such regular orbits. Malhotra (1995) concludes as well that the fact that some objects in the Edgeworth-Kuiper belt are in a (3:2) orbit-orbit resonance with Neptune, could be evidence of evolution of planetary orbits in the early Solar System.

5.3. A potential explanation

So why do we see relatively more period ratios between successive secondaries preferentially closer to irreducible Fibonacci fractions? Two aspects should be considered.

The first one is self-organization. In two recent papers, Aschwanden (2017, 2018) has addressed a possible physical origin for the occurrence of commensurabilities or near-commensurabilities of mean motions of successive secondaries, namely self-organization, by which spontaneous order can be created from an initial chaotic state. Self-organization is a property of dissipative nonlinear processes that are governed by an internal driver (here, gravitation) and a positive feedback mechanism (here, harmonic orbit resonances, Aschwanden, 2018). Dissipation could have occurred during early stages of protoplanetary or protosatellite nebulae while systems were in formation (see e.g. Tsiganis et al., 2005; Batygin et al., 2015; and Tamayo et al., 2017 for the Trappist-1 system) or by tidal transfer from the primary (Peale, 1976). Self-organization processes are found in astrophysics (Aschwanden, 2017; Aschwanden et al., 2018) and in many other fields (Aschwanden, 2018; Marov and Kolesnichenko, 2013; Haken, 2006; Feltz et al., 2006; Mishra et al., 1994; Gor'kavyi and Fridman,1991; Krinsky, 1984; Kernbach, 2008; Pontes, 2016).

The second aspect is linked to the preferential grouping of period ratios close to irreducible fractions of the type $\frac{p}{p+q}$ where $p$ and $q$ are small positive integers. The word "small" is important as the order of resonance or (quasi-) commensurability is given by the integer $q$ (i.e. the difference between denominator and numerator), while the intensity of the gravitational disturbance by a perturbing secondary on another smaller secondary is larger for small values of first, the order $q$, and second, the integer $p$. The main resonances that are observed (see e.g. Lemaitre, 2010; Veras and Ford, 2012; Aschwanden and Scholkmann, 2017) are $\frac{P_1}{P_2} = \frac{1}{2}, \frac{2}{3}, \frac{1}{3}, \frac{3}{5}, \frac{2}{5}$, for respectively $(q, p) = (1,1), (1,2), (2,1), (2,3), (3,2)$ (see Table 10).

TABLE 10: MAIN RESONANCES FOR $1 \leq q \leq 3$ AND $1 \leq p \leq 3$

| $\frac{P_1}{P_2} = \frac{p}{p+q}$ | $p = 1$ | $p = 2$ | $p = 3$ |
|---|---|---|---|
| $q = 1$ | $\frac{P_1}{P_2} = \frac{1}{2}$ | $\frac{P_1}{P_2} = \frac{2}{3}$ | $\left(\frac{P_1}{P_2} = \frac{3}{4}\right)$ |
| $q = 2$ | $\frac{P_1}{P_2} = \frac{1}{3}$ | -- | $\frac{P_1}{P_2} = \frac{3}{5}$ |
| $q = 3$ | $\left(\frac{P_1}{P_2} = \frac{1}{4}\right)$ | $\frac{P_1}{P_2} = \frac{2}{5}$ | -- |
| $q = 4$ | $\left(\frac{P_1}{P_2} = \frac{1}{5}\right)$ | -- | $\left(\frac{P_1}{P_2} = \frac{3}{7}\right)$ |

Note: Other resonances between parentheses.

Let us consider a simple model and reasoning by assuming the following three body configuration: a central primary of mass $M$, and two secondaries with masses $m_1$ and $m_2$ orbiting the central primary on elliptical orbits with semi-major axes $a_1$ and $a_2$, with $M \gg m_2 \gg m_1$ and $a_2 > a_1$, where the indices 1 and 2 refer respectively to the smaller inner and the outer perturber secondaries (the case of an inner pertuber is treated in a similar fashion; the case of coorbital secondaries for $q = 0$ is treated differently, see Robutel and Souchay, 2010). Let us assume that the less massive inner secondary 1 does not affect the larger outer secondary 2. The total orbital energy of the inner secondary is negative and reads classically (Lemaitre, 2010)

$$E_{Tot} = -\frac{GM}{2a_1} - \frac{Gm_2}{r_2} \sum_{k=2}^{\infty} \left(\frac{r_1}{r_2}\right)^k P_k(\cos \alpha)$$

(2)

where the first term is the orbital energy (or the Hamiltonian) of the classical two body problem and the second term, the potential energy $V_{Pert}$ due to the perturbation exerted by the outer secondary, with $G$ the gravitational constant, $r_1$ and $r_2$ the norm of the position vectors of the secondaries 1 and 2 with respect to the massive primary centre, $P_k$ the spherical Legendre polynomials and $\alpha$ the angle between the position vectors $\vec{r_1}$ and $\vec{r_2}$. For further simplifications, let's assume that the perturbing secondary is on a circular orbit (i.e. its eccentricity is nil, $e_2 = 0$ and $r_2 = a_2$), that the inner secondary's orbit is such that its apoapsis is still within the orbit of the outer perturber secondary orbit, i.e.

$$a_1(1 + e_1) < a_2$$

(3)

and that both secondary's orbits are coplanar. Then, the maximum perturbation by the outer secondary is exerted when both secondaries are in conjunction and the inner secondary is at its apoapsis, yielding $r_1 = a_1(1 + e_1)$, $\alpha = 0$ and $P_k(\cos 0) = 1$ in (2), giving for the first two terms of the perturbation potential function

$$V_{Pert} \approx -\frac{Gm_2}{a_2}\left(\left(\frac{a_1}{a_2}\right)^2 (1 + e_1)^2 + \left(\frac{a_1}{a_2}\right)^3 (1 + e_1)^3 + \cdots\right)$$

(4)

and with Kepler's third law, $\frac{a_1}{a_2} = \left(\frac{P_1}{P_2}\right)^{\frac{2}{3}}$, the perturbation potential reads

$$V_{Pert} \approx -\frac{Gm_2}{a_2}\left(\frac{P_1}{P_2}\right)^{\frac{4}{3}} (1 + e_1)^2 \left(1 + \left(\frac{P_1}{P_2}\right)^{\frac{2}{3}} (1 + e_1) + \cdots\right)$$

(5)

For the period ratio being an exact irreducible fraction of small integers,

$$\frac{P_1}{P_2} = \frac{p}{p + q}$$

(6)

with $p, q \geq 1$, the maximum perturbation from the outer secondary is exerted on the inner secondary every $(p + q)$ orbits, and is strongest for $q = 1$ and $p = 1$, i.e. $\frac{P_1}{P_2} = \frac{1}{2}$, yielding that the maximum perturbation is exerted every two orbits of the inner secondary, when both secondaries are in conjunction with the inner one at its apoapsis.

More generally, there are $q$ conjunctions between two secondaries having their period ratio equal to a fraction (6) and, counting the angle $\theta'$ from the apoapsis of the inner secondary, these $q$ conjunctions are located at successive angles $\theta' = \frac{2i\pi}{q}$ for $i$ integer, $1 \leq i \leq q$.

With $r_1 = a_1\left(\frac{1-e_1^2}{1+e\cos\theta}\right)$ where $\theta$ is the true anomaly of the inner secondary, the angle counted from its periapsis, and $\theta = \theta' - \pi$, yielding $\cos\theta = -\cos\theta'$, the total perturbing potential at the $q$ conjunctions (where $\alpha = 0$) reads more generally

$$V_{Pert} = -\frac{Gm_2}{a_2}\sum_{k=2}^{\infty}\left(\frac{a_1}{a_2}\right)^k \left(\frac{1-e_1^2}{1-e_1\cos\theta'}\right)^k = -\frac{Gm_2}{a_2}\sum_{i=1}^{q}\sum_{k=2}^{\infty}\left(\frac{P_1}{P_2}\right)^{\frac{2k}{3}}\left(\frac{1-e_1^2}{1-e_1\cos\left(\frac{2i\pi}{q}\right)}\right)^k$$

(7)

If we consider the repetitive effect of the outer secondary perturbations every $(p + q)$ orbits of the inner secondary, for a number $N$ of the inner secondary orbits, there are a total of $\frac{Nq}{(p+q)}$ conjunctions, including $\frac{N}{(p+q)}$ conjunctions with the inner secondary at its apoapsis, $\frac{N}{(p+q)}$ conjunctions with the inner secondary at its periapsis if $q$ is even, and $\frac{N q'}{(p+q)}$ lateral conjunctions (i.e. with the inner secondary not at its periapsis nor at its apoapsis) with $q' = (q - 2)$ if $q$ is even or $q' = (q - 1)$ if $q$ is odd.

Therefore, a simple way to express the long-term effect of the perturbation is to add up the repetitive perturbations of all conjunctions over $N$ orbits of the inner secondary. Taking $N = 1$ and multiplying the perturbation potential in (7) by $\frac{1}{(p+q)}$ yield generally, with (6)

$$\frac{V_{Pert}}{p+q} = -\frac{Gm_2}{a_2}\frac{1}{p+q}\sum_{i=1}^{q}\sum_{k=2}^{\infty}\left(\frac{p}{p+q}\right)^{\frac{2k}{3}}\left(\frac{1-e_1^2}{1-e_1\cos\left(\frac{2i\pi}{q}\right)}\right)^k$$

(7)

which for a first order resonance ($q = 1$) for example, reduces for the first three terms to

$$\frac{V_{Pert}}{p+q} \approx -\frac{Gm_2}{a_2}(1+e_1)^2\frac{1}{p+q}\left(\frac{p}{p+q}\right)^{\frac{4}{3}}\left(1+\left(\frac{p}{p+q}\right)^{\frac{2}{3}}(1+e_1)+\left(\frac{p}{p+q}\right)^{\frac{4}{3}}(1+e_1)^2+\cdots\right)$$

(8)

From (7), we form the ratio $\varepsilon = \frac{-V_{Pert}}{(p+q)(Gm_2/a_2)}$, called the resonance relative repetitive perturbation. Figure 10 shows the distribution of the values of $\varepsilon$ for $e_1 = 0.1$ and for $a_1 < 0.8\, a_2$ from condition (3). Note that the values of $\varepsilon$ are relatively insensitive to the values of $e_1$ for $0 \leq e_1 \leq 1$, and that this simplified model cannot be used for orbital configurations where the values of $a_1$ and $a_2$ are to close.

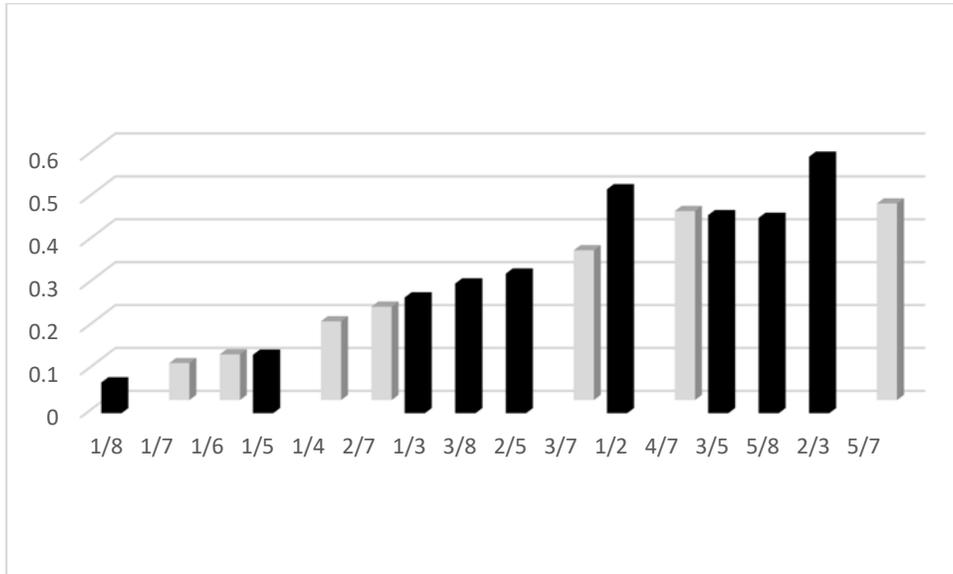

Figure 10: Distribution of values of the resonance relative repetitive perturbation $\varepsilon$ for resonance ratios up to $\frac{5}{7}$ (black: Fibonacci fractions; grey: other fractions).

This simple model and analysis show that (1) the main resonance ratios $\frac{P_1}{P_2} = \frac{1}{2}, \frac{2}{3}, \frac{1}{3}, \frac{3}{5}, \frac{2}{5}$ of order 1, 2 and 3 are stronger than other resonances, (2) these ratios correspond to Fibonacci fractions, (3) other resonance ratios correspond to weaker resonances and appear less often, in particular the resonance relative repetitive perturbation ε of the two weaker resonance ratios $\frac{P_1}{P_2} = \frac{4}{7}, \frac{3}{7}$ of order 3 and 4 (whose bins are adjacent to the bin of $\frac{1}{2}$, see Figure 1), are lower than those of adjacent $\frac{P_1}{P_2} = \frac{2}{5}, \frac{1}{2}$ and $\frac{3}{5}$.

Therefore, it can be concluded that there is a natural preference for period ratios of consecutive secondaries to be close to irreducible fractions of Fibonacci numbers, although exceptions exist, e.g. exoplanetary systems Kepler-444, 55 Cnc, HD 219134, Kepler-296, Kepler-62 (see Table 5). Despite the fact that Fibonacci numbers are found extensively in other fields of science and technology (Koshy, 2001; Pletser, 2017c) in systems and configurations exhibiting a minimum of energy, the above simple model shows that the preference for period ratios of successive secondaries to be close to irreducible fractions of Fibonacci numbers is associated to stronger resonances or (quasi-)commensurabilities with repeated orbital perturbations. Furthermore, the analysis of Section 4 for different sets of minor planets show that the tendency for period ratios of successive secondaries to be associated to fractions of Fibonacci numbers is stronger for more "regular" sets of minor planets, i.e. on less inclined and less elliptical orbits. So, orbital regularity can be associated to quasi-commensurabilities with period ratios close to fractions of Fibonacci numbers, or to a natural evolution of secondaries toward more "regular" (less inclined and less elliptical) orbits.

However, one must be cautious with interpretations that can be made with Fibonacci numbers, as none of the specific characteristics of Fibonacci numbers, mainly that every Fibonacci number is the sum of the two previous ones, is apparent in orbital period ratios of secondaries or other orbital characteristics. Period ratios of successive secondaries are observed to be preferentially closer to irreducible fractions of small integers of the form $\frac{p}{p+q}$, most of them being formed by the second to the sixth Fibonacci integers.

## 6. Conclusions

It was shown that orbital period ratios of successive secondaries in the Solar planetary and giant satellite systems and in exoplanetary systems are preferentially and significantly closer to irreducible fractions formed with the second to the sixth Fibonacci numbers (between 1 and 8) than to other fractions, in a ratio of approximately 60% vs 40%, although there are less irreducible fractions formed with Fibonacci integers between 1 and 8 than other fractions.

Furthermore, if sets of minor planets are chosen with gradually smaller inclinations and eccentricities, one observes that the proximity to Fibonacci fractions of their period ratios with Jupiter or Mars' period tends to increase for more "regular" sets with minor planets on less eccentric and less inclined orbits. Therefore, orbital period ratios closer to Fibonacci fractions could indicate a greater regularity in the system.

Finally, a simple model could explain why the resonances of the form $\frac{P_1}{P_2} = \frac{p}{p+q}$ with small values of integers $p$ and $q$ are stronger and more commonly observed. However, one should be cautious also with potential interpretation as the specific characteristics of Fibonacci numbers, i.e. every Fibonacci number is the sum of the two previous numbers, is not observed in orbital period ratios of successive secondaries.

**Acknowledgements**

This research has made use of data and/or services provided by the International Astronomical Union's Minor Planet Center. Prof. D. Huylebrouck and Prof. L. Basano kindly provided comments on early versions of the manuscript. Stimulating discussions with C. Ducrest are also acknowledged.